# Atom Probe Tomography as an Emerging Tool for Understanding Defect-driven Mechanisms in $HfO_2$-based Ferroelectrics

Kasper Hunnestad[1,2*], Catherine Dubourdieu[3,4*], Alexei Gruverman[5*], Dennis Meier[1,6,7*]

[1]Department of Materials Science, Norwegian University of Science and Technology (NTNU), Trondheim, Norway

[2]Department of Electronic Systems, Norwegian University of Science and Technology (NTNU), Trondheim, Norway

[3]Helmholtz-Zentrum Berlin für Materialien und Energie, Berlin, Germany

[4]Physical and Theoretical Chemistry, Freie Universität Berlin, Berlin, Germany

[5]Department of Physics and Astronomy, University of Nebraska, Lincoln, NE, USA

[6]Faculty of Physics and Center for Nanointegration Duisburg-Essen (CENIDE), University of Duisburg-Essen, Duisburg, Germany

[7]Research Center Future Energy Materials and Systems, Research Alliance Ruhr, 44780 Bochum, Germany

*corresponding authors

**Abstract**

$HfO_2$-based ferroelectrics are essential for the next generation of CMOS-compatible memory and logic devices, yet their performance is governed by a complex interplay between oxygen vacancies, dopants, and structural defects that remains an active area of investigation. These defects shape the function-critical dynamic phenomena, such as polar phase stabilization, wake-up, fatigue, and imprint. In this Perspective, we review the limitations of established high-resolution structural characterization techniques and propose atom probe tomography (APT) as a powerful tool for the three-dimensional (3D), atomic-scale mapping of all constituent species in hafnia-based ferroelectric systems. By resolving individual dopants, vacancy clustering, and interfacial segregation, APT can facilitate a quantitative understanding of defect-property relations in hafnia-based ferroelectrics. We discuss current experimental challenges for APT application to ferroelectric oxides, demonstrate a proof-of-concept of atomic-scale reconstruction in a hafnia-based device stack, and highlight the potential of APT to guide the development of ferroelectric structures with enhanced reliability and performance.

## 1. Introduction

$HfO_2$-based compounds are a class of complementary metal–oxide–semiconductor (CMOS)-compatible, high-dielectric-constant materials that have become essential dielectrics for nanoelectronics. By replacing conventional $SiO_2$ in field-effect transistors, they help suppress leakage currents and enable continued device scaling toward smaller, more energy-efficient electronics[1]. Additional application opportunities emerged following the discovery of ferroelectricity in silicon-doped $HfO_2$[2–4], which triggered world-wide interest in the ferroelectric fluorite-type systems and their integration in ferroelectric field effect transistors (FeFET) and random access memory (FeRAM)[5].

Since the first reports on ferroelectricity in $HfO_2$[2], substantial progress has been made in the synthesis, experimental characterization, and understanding of the physical properties of this family of materials[6]. Examples include stabilization of the ferroelectric phase through defect chemistry, doping and strain engineering, observation of domain kinetics, delineation of the mechanisms of dynamic phenomena[7–9], and the engineering of thin films with enhanced functional responses[10]. However, device concepts that exploit ferroelectric order are still under intensive investigation and have yet to be fully realized [11–13]. This challenge is closely related to the structural complexity of this seemingly simple binary oxide and the fundamental impact atomic defects have on its physical responses. For instance, it is now established that $HfO_2$-based ferroelectrics are highly sensitive to oxygen vacancies, which play a crucial role in phase stability and strongly influence the switching behavior under applied electric fields (**Figure 1**).

In general, a plethora of defects arising in doped hafnia systems can enhance or diminish the desired functionality[14–17]. Thus, to understand and control defect-related behavior, there is a strong need for adequate characterization techniques from the macroscopic to the atomic level that can capture the different point defects and their spatial distribution, including interfacial accumulation, cluster formation, and gradients. This need is reflected by the large variety of advanced experimental methods that have been adopted, as well as in the continued development of new techniques to address the rich defect physics of this intriguing class of materials[18].

In this Perspective, after a brief overview of the characterization techniques that focus on defect-driven properties in hafnia-based ferroelectrics, we highlight a method with significant untapped potential for studying these materials: atom probe tomography (APT). Drawing analogies with other ferroelectric systems, we examine the emerging opportunities this technique offers and how it can deepen our understanding of defect-driven mechanisms in $HfO_2$-based ferroelectrics.

## 2. Defect-driven effects in $HfO_2$-based ferroelectrics

The stabilization of the non-centrosymmetric orthorhombic $Pca2_1$ phase that enables ferroelectricity in hafnia-based films remains one of the most important topics in the field. Various dopants, such as Zr, Si, Al, Y, and La, have been introduced to favor this phase over the monoclinic and tetragonal phases[19]. The dopant atoms alter both the lattice strain and local chemistry, influencing the free energy landscape among the various phases. For instance, Zr substitution tends to reduce the energy difference between the tetragonal and orthorhombic phases, whereas Si or Al doping stabilizes ferroelectricity primarily through lattice mismatch and associated local stress fields[10,19,20]. However, the precise mechanisms of stabilization are not fully understood, in part because individual dopants are difficult to resolve experimentally. As a result, much of the related research relies on advanced modeling and computational approaches, including density functional theory, molecular dynamics simulations, and phase-field modeling, to determine defect formation pathways and energetics[21]. Still, the dynamic and kinetic processes, such as domain switching, often lie beyond the scope of conventional modelling, while the complexities of real systems with polycrystalline structures and multi-defect interactions remain challenging to capture.

Going beyond the stabilization of the ferroelectric phase, point defects are essential for the electric-field response and the polarization switching behavior[22]. Similar to conventional ferroelectrics, such as lead zirconate titanate ($Pb(Zr_xTi_{1-x})O_3$) and barium titanate ($BaTiO_3$), the electrical reliability of $HfO_2$-based ferroelectrics is strongly shaped by different dynamic phenomena - that is, wake-up, fatigue, and imprint effects - each governed by distinct microscopic mechanisms[22–24]. In many cases, pristine $HfO_2$-based films exhibit a weak polarization response, showing constricted or asymmetric hysteresis loops (**Figure 1**). Upon repeated bipolar cycling, the remanent polarization increases, reflecting a substantial wake-up effect. This effect can result from the field-induced structural transition to the polar phase, converting nonpolar or weakly polar regions into the ferroelectric orthorhombic phase, or from the redistribution of oxygen vacancies and release of the trapped charges that initially impede domain switching[22].

This structural phase transition is facilitated by local field enhancement and defect-assisted stress relaxation, effectively increasing the volume fraction of the ferroelectric phase. It was recently proposed that the change of the longitudinal piezoelectric coefficient - from positive to negative values - upon electrical cycling in $(Hf,Zr)O_2$ capacitors resulted from a phase transformation of $(Hf,Zr)O_2$ from a weakly-developed to well-developed orthorhombic phase and that it could originate from both ionic and vacancy motions[25]. Additionally, depending on the chemical composition, field-induced ferroelastic domain switching can play a prominent role in the wake-up mechanism due to the disorder-driven domain nucleation and nonlinear domain wall dynamics, highlighting the importance of the difficult-to-grasp oxygen vacancies, dopants, and point defects in general. In addition, microstructural pre-conditions for the wake-up effect, such as grain size, interface structure and chemical composition are not yet fully controlled.

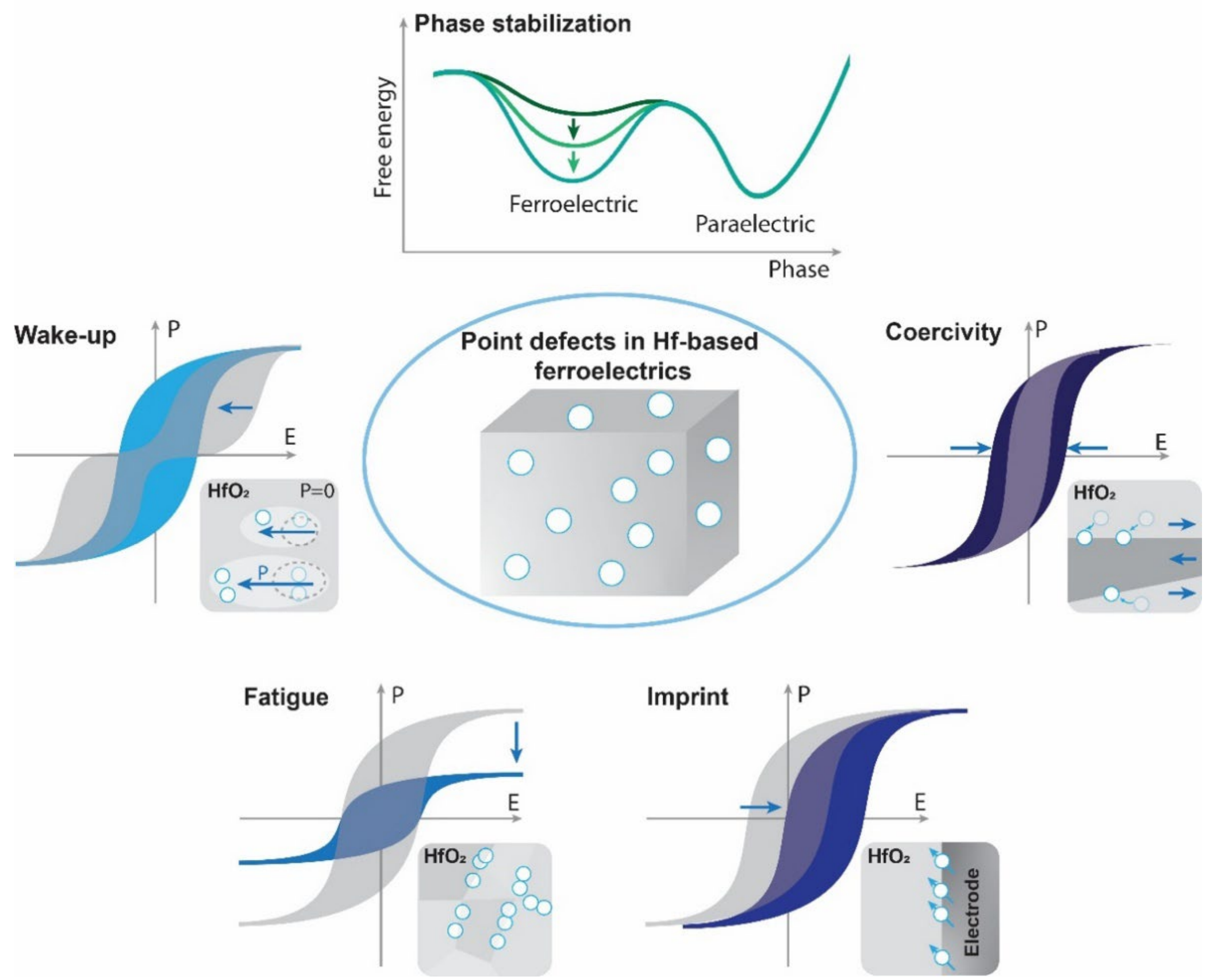

**Figure 1**: Impact of point defects on physical properties of ferroelectric hafnia-based films. Stabilization of the ferroelectric phase is driven both by dopants and the mobile defect redistribution during switching. Excessive defect migration and segregation, however, can also lead to detrimental effects, such as fatigue, imprint and increased coercivity.

It has been observed that excessive cycling can again reverse the wake-up related polarization increase, as defect migration and interfacial degradation eventually trigger fatigue and imprint. For example, development of the oxygen-deficient regions at the TiN/$HfO_2$ interface results in charge entrapment that locally screens the polarization, reducing the effective switchable volume[11]. Moreover, the strong coupling between electric-field cycling and defect motion introduces electrochemical asymmetry: electrons and oxygen vacancies migrate in opposite directions, promoting local redox reactions. Asymmetric accumulation of oxygen vacancies or trapped charges at the interfaces create an internal bias resulting in imprint development (**Figure 1**)[26,27]. The coupling between imprint and fatigue forms a self-reinforcing cycle: imprint introduces field asymmetry, causing incomplete or unbalanced switching; this in turn enhances local defect generation, accelerating further fatigue. Experimental studies of (Hf,Zr)$O_2$ capacitors reveal co-evolving loop constriction and horizontal shift after $10^5$-$10^8$ cycles, confirming their common defect-mediated origin[24]. A full understanding of this dynamical behavior requires mapping of oxygen vacancies and their redistribution during switching, a challenge that has motivated several pioneering microscopy studies[8,28,29]. The realization of quantitative measurements in three-dimensional (3D) space, however, remains an unsolved issue.

The complexity of defect-mediated phenomena generally increases in device architectures. As in other ferroelectrics, the properties of $HfO_2$-based structures depend critically on their interfaces with electrodes. These interfaces can strongly shape the material response through built-in electric fields, interfacial dead layers, chemical activity, and lattice-mismatch strain[30]. The work-function difference between the metal electrode and the hafnia layers establishes built-in fields that can either stabilize or suppress one of the polarization states. Electrodes such as TiN, W, Ru, and Pt differ markedly in their screening capabilities and chemical stability, which in turn affects coercivity and imprint[31]. Moreover, the formation of interfacial oxide layers (e.g., $TiO_x$ at TiN/$HfO_2$) and dipoles modifies local electric potentials[8] and can introduce low-κ dead layers that reduce the effective switchable polarization. Chemical reactions and diffusion across the film-electrode interfaces play an important role in hafnia device reliability. For example, nitrogen out-diffusion from TiN electrode layers alters oxygen-vacancy concentrations in $(Hf,Zr)O_2$ capacitors, modifying both leakage and polarization[32,33].

All these examples clearly show that progress in $HfO_2$-based ferroelectrics requires a detailed understanding of defect physics, and that their sensitivity to atomic-scale defect chemistry is exceptionally complex compared with other ferroelectrics. Oxygen vacancies, especially, play a paradoxical role in hafnia ferroelectrics. On the one hand, at moderate concentrations, they assist in stabilizing the ferroelectric orthorhombic phase by lowering the energy barrier for its formation and enabling easier domain wall motion[7]. As mentioned above, their redistribution during early cycling also contributes to the wake-up effect[34]. On the other hand, an excessive oxygen-vacancy concentration introduces significant drawbacks. Accumulation of oxygen vacancies near electrodes creates an internal bias that leads to imprint and asymmetric switching[35]. Over time, vacancy drift and clustering contribute to ferroelectric ageing, enhanced leakage, and retention loss[36]. Moreover, vacancy migration pathways can serve as conducting channels, compromising ferroelectric switching and electrical endurance. Beyond independent oxygen vacancies, the interplay with the dopants and structural defects remains largely unexplored.

## 3. Observing the property-defining point defects

Given the crucial role of point defects for the physical properties of $HfO_2$-based ferroelectrics, it is not surprising that extensive efforts have been devoted to characterizing these defects and developing suitable methods to probe them. Analysis of oxygen vacancies is particularly challenging at the local scale, despite their substantial impact being readily detectable in macroscopic measurements. Consequently, a range of advanced microscopy and spectroscopy techniques has been employed to investigate their associated effects (**Figure 2**)[18]. Macroscopically, the impact of oxygen vacancies can be deduced from the switching measurements by a positive-up-negative-down (PUND) method[37,38]. Furthermore, defect-driven changes in bonding chemistry can be resolved via spectroscopic techniques, such as X-ray photoelectron spectroscopy (XPS)[30]. Insights from XPS have proven useful for studying electrode interface chemistry, particularly in relation to the applied growth conditions. For example, Müller *et al*. showed from XPS data a $TiO_2$ intralayer that formed at the interface between $(Hf,Zr)O_2$ and the TiN electrode, blocking diffusion of oxygen across[30].

Electron microscopy has been widely used to map the spatial distribution of defects and their effects. Electron energy loss spectroscopy (EELS) and energy-dispersive X-ray spectroscopy (EDX) can probe oxygen defects and assess local stoichiometry with high sensitivity, which has been utilized for visualizing the oxygen vacancies before and after a.c. cycling of $(Hf,Zr)O_2$ capacitors, revealing their non-homogenous distribution[36,39]. Despite the outstanding opportunities that EELS offers for chemical analysis at the atomic scale, the technique is inherently limited to two-dimensional (2D) projections and relies heavily on model-based interpretation. Other electron-based imaging approaches, such as electron holography, enable direct imaging of local electrostatic potentials and interface charges inside the material, giving additional valuable insights into the dynamical aspects related to polarization[8]. Integrated differential phase contrast scanning transmission electron microscopy (iDPC-STEM) is particularly powerful when it comes to direct imaging of light elements, such as oxygen, offering the possibility to visualize columns of oxygen vacancies and associated changes in the crystal structure. Individual oxygen vacancy columns can now be directly imaged. In a pioneering study, this capability revealed that oxygen migration through the $(Hf,Zr)O_2$ layer is closely linked to ferroelectric switching.[29] However, in addition to its restriction to 2D-projections and local unit cell orientation, statistical uncertainties remain a challenge when probing oxygen defects by iDPC-STEM.

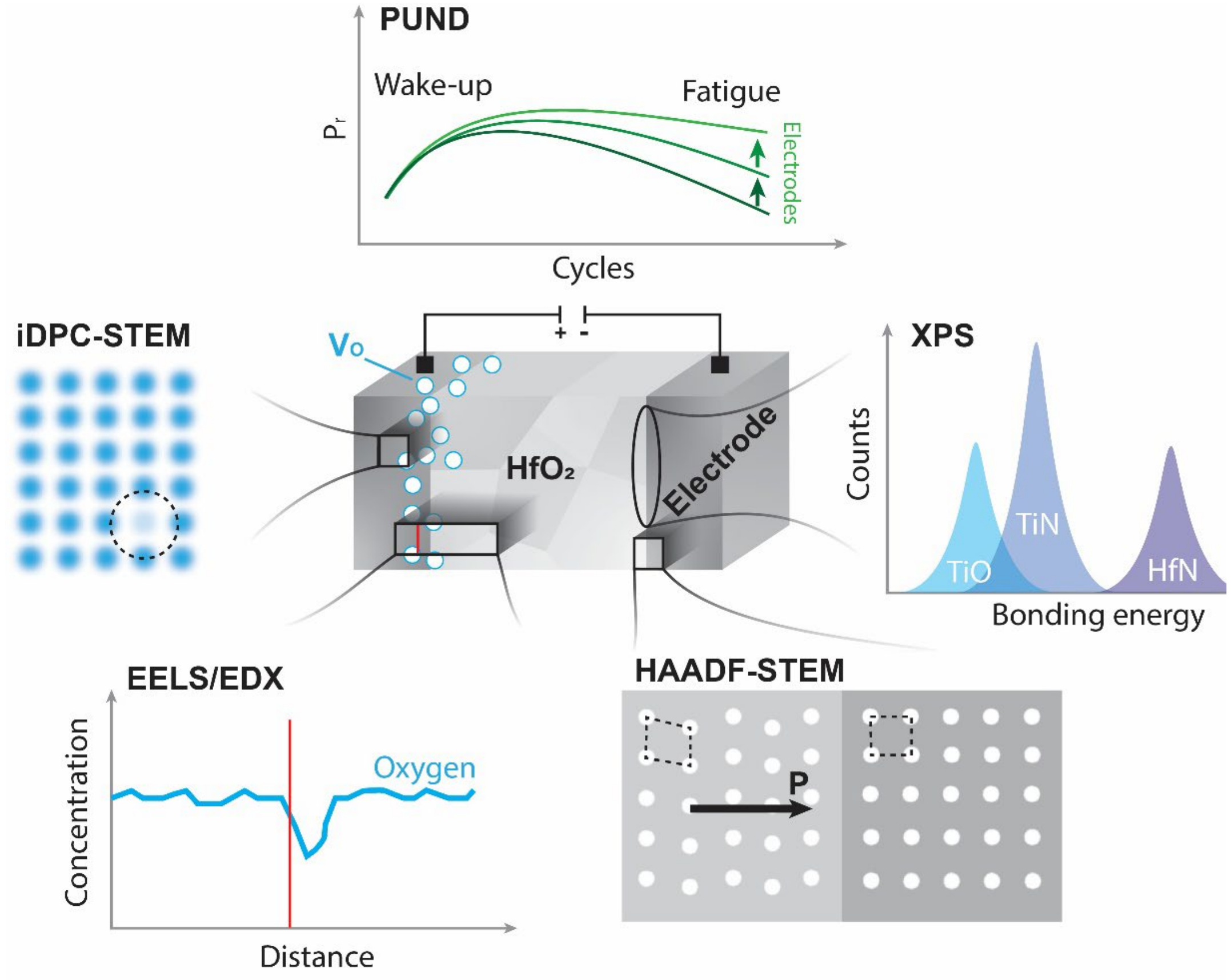

**Figure 2:** A wide range of microscopy and spectroscopy techniques has been applied to investigate $HfO_2$ systems. Macroscopic polarization values and their evolution as a function of a.c. cycling are measured using PUND, a switching-based method that averages over the entire sample. Bonding states and interface chemistry are probed by XPS, which is surface-sensitive and localized to planar regions. The atomic structure at interfaces and the resulting polarization directions can be resolved from the cation distribution using HAADF-STEM imaging, whereas iDPC enables visualization of lighter elements; however, both techniques are limited to specific zone axes and provide two-dimensional projections. Compositional information obtained by EELS and EDX is essential for mapping defect segregation.

Hence, there are plenty of defect-related fundamental questions to be answered that elude the conventional microscopy approaches. Low dopant or defect concentrations, for instance, can have substantial impact on the system performance, but they are often not detected in imaging experiments due to a lack of sensitivity. Atomic-level interactions between the defects, including clustering and the formation of defect-pairs, require full 3D imaging capabilities with atomic-scale resolution, and they are often too computationally demanding for numerical approaches. Furthermore, continuous scaling of device structures makes them increasingly sensitive to local structural and chemical inhomogeneities. This challenge is further amplified in three-dimensional multilayer architectures, which must remain reliable under sustained a.c. cycling. Characterization techniques capable of resolving defects that lead to electrical malfunction are therefore essential. Thus, there is a clear need for additional microscopy approaches to tackle such defect-chemistry questions and challenges, including 3D aspects and quantifiable data.

## 4. Atom probe tomography and investigation of hafnia-based ferroelectrics

A technique that has the potential to fill some of the experimental gaps associated with the characterization of oxygen defects in $HfO_2$ and their impact on the functional properties is atom probe tomography (APT)[40]. APT is well established in metallurgy and semiconductor research where it is used for studying compositional heterogeneity[41,42]. Interfaces and defects, including grain boundaries and dislocations, are particularly important targets as the high spatial resolution and chemical sensitivity of APT can provide otherwise inaccessible insights. Probably, one of the most important study cases is precipitation in semiconductors where APT enables both direct imaging and statistical analysis of precipitation from early-stage dopant distribution to large-scale clustering phenomena. Furthermore, as time-of-flight spectroscopy is at the heart of APT, isotopic analysis is possible, which is relevant to nuclear material research and geochronology[43,44].

For APT experiments, the samples are prepared in the form of a needle with an apex curvature radius of about 50 nm. Under an electric potential of around 5 kV, this geometry sustains an electric field strong enough to induce field evaporation of the atoms from the specimen surface as sketched in **Figure 3**[45]. The ions follow the electrostatic trajectories defined by the specimen itself, eliminating the need for lenses, and are detected by a delay-line detector. By collecting the evaporated atoms, while measuring the time-of-flight and impact position, a 3D reconstruction of the atomic distribution can be achieved with near-atomic resolution[40,46]. It is important to note that the resulting reconstruction is inherently built by millions of individual atoms, accessible completely independent of each other. This is one of the primary reasons why APT is uniquely positioned to resolve fine clustering, as these effects tend to vanish in a 2D projection. Thus, resolving individual dopants[47], defects clusters and their segregation to interfaces, such as domain or grain boundaries, becomes possible[48,49]. In addition, APT can, in principle, detect all species present in the material with equal sensitivity, including light elements, such as hydrogen. This capability is essential for studying phenomena such as embrittlement as it reduces the risk of missing essential information and systematic biases when considering only a single species[50,51].

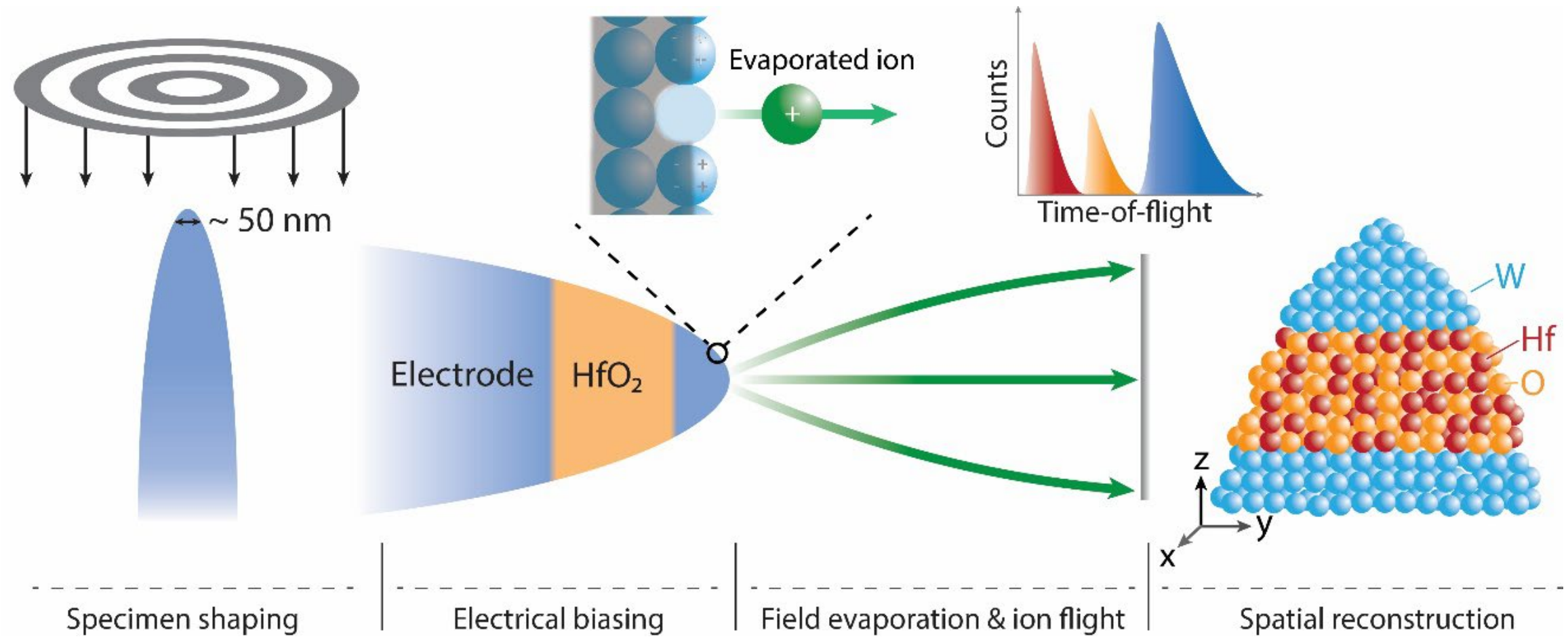

**Figure 3:** Schematics of APT analysis. First, a bulk sample is shaped into a nanoscale needle, with a tip radius of about 50 nm. After applying an electrical bias, surface atoms are ionized and evaporated from the specimen and recorded at the detector. Using the time-of-flight and trajectory, a 3D reconstruction of the atomic distribution can be made.

In contrast to metals, the application of APT to complex oxides has only recently begun to attract broader attention. This slower uptake is partly due to material-specific challenges, including specimen fracture associated with the brittleness of these materials. Optimization of analysis conditions, rigorous material screening, and technical advances in successive generations of APT have gradually reduced these experimental challenges. As a result, the broader potential of atomic-scale defect analysis in ferroelectrics, and in complex oxides more generally, has now been demonstrated. An important step concerning the application of APT to ferroelectrics was the investigation of $Pb(Zr,Ti)O_3$ in 2014[52]. Here, synthesis-related Ti and Zr clustering was observed beyond conventional TEM investigations, introducing APT as a powerful tool for atomic structure analysis in ferroelectrics. About a decade later, correlations between polarization discontinuities and chemical segregation were investigated[49,53,54]. For example, APT studies of ferroelectric $(LuFeO_3)_9/(LuFe_2O_4)_1$ superlattices revealed the segregation of oxygen vacancies into the $LuFe_2O_4$ monolayers, forming nanometer-sized clusters within individual atomic planes[54]. Importantly, a direct correlation to the polar order was observed, where oxygen vacancies were found to be responsible for stabilizing the charged head-to-head and tail-to-tail domain walls by redistributing mobile electrons. These studies reflect the remarkable sensitivity and potential of APT when it comes to the imaging of individual point defects, and especially in regard to their interaction with interfaces.

For hafnia-based systems, APT can provide insight into whether phase stabilization arises from homogeneous doping or local clustering. It can also help clarify how strain and interface energies interplay with dopant gradients and oxygen non-stoichiometry in determining phase stability. By simultaneously detecting all elements with high sensitivity and spatial resolution, APT has the capability to disentangle the coupled roles of dopants, defects, and interfaces that govern the complex switching behavior of hafnia-based ferroelectrics. Furthermore, resolving the nanoscale compositional gradients by APT could enable quantitative mapping of interfacial species diffusion. With direct visualization of the defect migration and interfacial diffusion before and after a.c. cycling, it would be possible to distinguish reversible processes that underline the wake-up effects from irreversible damage leading to fatigue. Such correlation will be critical to design defect-tolerant architectures with extended endurance. The nanoscale insight into the defect chemistry provides a pathway toward rational defect engineering, enabling predictive synthesis of stable and durable high-performance ferroelectric architectures in hafnia-based devices.

As oxides are prone to fracture, one of the main challenges in utilizing APT is sample preparations and optimizing the analysis conditions. Although APT has been used in $HfO_2$-based high-κ/metal gate stacks to correlate interdiffusion with the threshold voltage of the FETs[55], $HfO_2$ thin films with ferroelectric properties remain untested. Data acquired from a W/(Hf,Zr)$O_2$/W device stack (**Figure 4a**) serve as a proof-of-concept and demonstration of the feasibility of using APT to study $HfO_2$-based ferroelectrics. The entire device structure was fabricated into an APT specimen using a focused ion beam (FIB) (**Figure 4b**). By first extracting a larger region from the material, followed by sequential top-down milling along the needle axis, the specimen was thinned down to around 100 nm in diameter[56]. Formation of amorphous surface layers and Ga implantation was avoided by using low-energy irradiation, similar to sample preparation for STEM[57]. In case of a thin film (or a multi-layer system), it is preferable to align the film with the needle axis to avoid fracture. Before the APT analysis of the (Hf,Zr)$O_2$ stack shown in **Figure 4**, the sample was placed in ultra-high vacuum and cooled to 25 K. Afterwards, the 3D structure of the film was reconstructed from the measured atomic distribution (**Figure 4c**). The advantage of such 3D data acquisition approach is complete flexibility in post-experiment analysis. This is illustrated in **Figure 4d**, where the specimen image is rotated such that it can be viewed from different directions providing an insight into the spatial distribution of structural and chemical inhomogeneities.

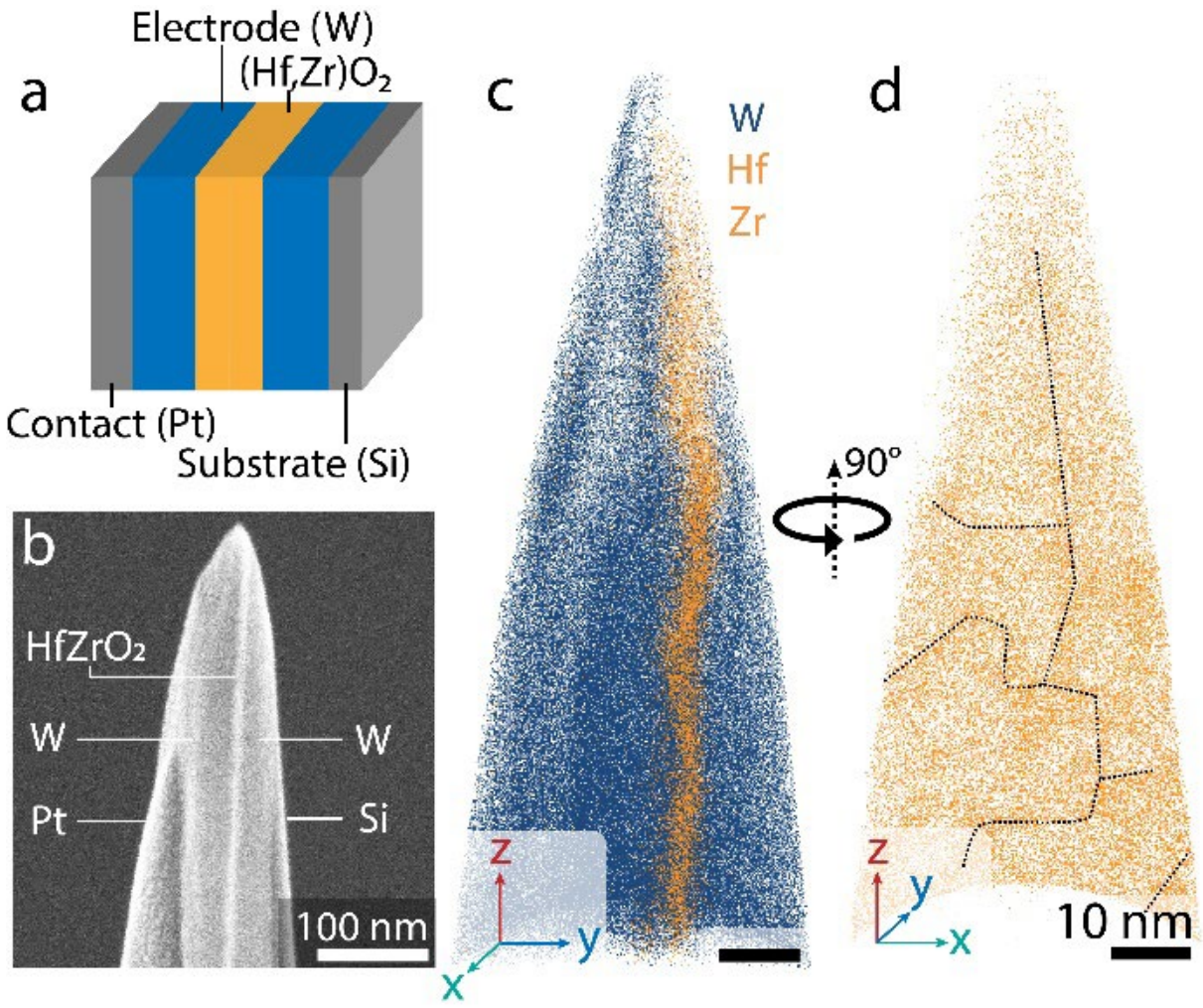

**Figure 4:** Proof-of-concept experiment. (a) Schematic representation of a W/$(Hf,Zr)O_2$/W device stack. (b) SEM image of the APT specimen prepared from the $(Hf,Zr)O_2$ stack. (c) 3D APT reconstruction of the specimen, centered around the $(Hf,Zr)O_2$ thin film. The $(Hf,Zr)O_2$ thin film atoms are all colored orange for visibility. The apparent gradual expansion of the thin film from top to bottom in the reconstruction, is one example of the reconstruction artefacts caused by differences in field evaporation rates between W and $(Hf,Zr)O_2$. (d) Side-view of the $(Hf,Zr)O_2$ thin film, showing various inhomogeneities, potentially grain boundaries, appearing as black lines across the sample.

Despite the advantages of APT, there are practical challenges to consider. There is a substantial difference in field evaporation strength between metallic electrodes and hafnium oxide, promoting fracturing during analysis and substantial reconstruction artefacts, obscuring the difference between real and artificial chemical anomalies. This limits the ability to acquire statistically robust data and highlights the need for further optimization of sample preparation and analysis conditions. However, recent developments in APT can be applied to mitigate this problem. Next-generation APT instruments have already shown a dramatic improvement in handling complex oxide materials, both in terms of fracture rate and field evaporation artefacts[35]. Novel advanced sample preparation techniques, or different electrode materials, can also be used to further increase the survival rate[58].

## 5. Outlook

The interplay between oxygen vacancies, dopants, structural defects, and dynamic phenomena remains one of the most complex and debated aspects of $HfO_2$-based ferroelectrics, fundamentally limiting their deployment in memory and logic devices. Atom probe tomography (APT) offers a powerful route to address this challenge by providing 3D atomic-scale mapping of all constituent elements. APT offers direct access to the oxygen non-stoichiometry and subtle compositional gradients across film thicknesses and interfaces and provide information that is inaccessible by 2D projection-based techniques. This opens the door to disentangling the role of defect chemistry in the ferroelectric behavior, distinguishing whether device wake-up, fatigue, and imprint are driven by vacancy redistribution, interfacial redox processes, or electrode-driven species diffusion.

At the same time, applying APT to hafnia-based ferroelectrics also presents experimental challenges. Hafnia specimens are prone to fracture under the intense electrostatic pressures required for field evaporation. Differences in evaporation fields between metallic electrodes and hafnia layers can cause fractures and severe reconstruction artifacts, obscuring real chemical features. First proof-of-concept experiments, however, demonstrate the general feasibility and potential of APT-based investigations.

Looking forward, the technical challenges are expected to be solved in the near-future and be outweighed by the benefits provided by combining APT with high-resolution electron microscopy methods. We believe that this combination of advanced microscopy approaches will facilitate a much deeper understanding of the physical mechanisms behind the unique physical properties of hafnia-based ferroelectrics.

## Acknowledgements

The authors are grateful to Constantinos A. Hatzoglou for his support with the APT lab facilities. M. Holzer, M. Hamid Raza and Dong-Jik Kim (HZB) are acknowledged for preparing the samples for APT analysis. C.D. acknowledges funding from the European Union's Horizon research and innovation programme under grant agreement 101135398 (FIXIT). Views and opinions expressed are, however, those of the authors only and do not necessarily reflect those of the European Union or European Commission. The Research Council of Norway (RCN) is acknowledged for its support to the Norwegian Micro- and Nano-Fabrication Facility, NorFab, project number 295864, and the Norwegian Laboratory for Mineral and Materials Characterization, MiMaC, project number 269842/F50. K.A.H. and D.M. thank the Department of Materials Science and Engineering at NTNU for direct financial support. A.G. acknowledges support by the National Science Foundation (grant DMR-2419172).